\documentclass[]{emulateapj}

\shorttitle{Fluorescent Ly$\alpha$ emitters}
\shortauthors{Cantalupo, Lilly \& Porciani}
\slugcomment{The Astrophysical Journal, 657:135-144, 2007 March 1}

\newcommand{\ergscm}{$\mathrm{erg}\ \mathrm{s}^{-1}\mathrm{cm}^{-2}$}
\newcommand{\ergscmarcsec}{$\mathrm{erg}\ \mathrm{s}^{-1}\mathrm{cm}^{-2}\mathrm{arcsec}^{-2}$}
\newcommand{\ergscmAA}{$\mathrm{erg}\ \mathrm{s}^{-1}\mathrm{cm}^{-2}\mathrm{\AA}^{-1}$}
\newcommand{\plus}{$\ \ +\ \ \ $}
\newcommand{\none}{$\ \ \ \ \ \ \ $}
\newcommand{\minus}{$\ \ -\ \ \ $}

\begin{document}
\def\simlt{\mathrel{\rlap{\lower 3pt\hbox{$\sim$}} \raise
        2.0pt\hbox{$<$}}} \def\simgt{\mathrel{\rlap{\lower
        3pt\hbox{$\sim$}} \raise 2.0pt\hbox{$>$}}}
\def\simgt{\mathrel{\rlap{\lower 3pt\hbox{$\sim$}} \raise
        2.0pt\hbox{$>$}}} \def\simgt{\mathrel{\rlap{\lower
        3pt\hbox{$\sim$}} \raise 2.0pt\hbox{$>$}}}

\title{Plausible fluorescent Ly$\alpha$ emitters around the $z=3.1$ QSO0420-388\altaffilmark{*}}
  
\author{Sebastiano Cantalupo, Simon J. Lilly and Cristiano Porciani}
\affil{Institute for Astronomy, ETH Z\"urich, Zurich, Switzerland}

\altaffiltext{*}{Based on observations carried out at the European
Southern Observatory, Paranal, Chile, program 076.A-0837A.2}
   
\email{cantalupo@phys.ethz.ch}


\begin{abstract}
We report the results of a survey for fluorescent Ly$\alpha$ emission carried out in the
field surrounding the $z=3.1$ quasar QSO0420-388 using the Focal Reducer/Low Disperion Spectrograph 2 (FORS2) instrument on the Very Large Telescope (VLT). 
We first review the properties expected for fluorescent Ly$\alpha$ emitters, compared with
those of other non-fluorescent Ly$\alpha$ emitters.  
Our observational search detected 13 Ly$\alpha$ sources sparsely sampling a 
volume of $\sim14000$ comoving Mpc$^3$ around
the quasar. 
The properties of these in terms of (1) the line equivalent width, (2) the line
profile and (3) the value of the surface brightness related to the
distance from the quasar, all suggest that several of these may be plausibly
fluorescent. Moreover, their number is in good agreement with
the expectation from theoretical models.
One of the best candidates for fluorescence is sufficiently far behind 
QSO0420-388 that it would imply that the quasar has been active for (at least) $\sim$60 Myr.
 Further studies on such objects will give information
about proto-galactic clouds and  
on the radiative history (and beaming) of the high-redshift quasars.
\end{abstract}


\section{Introduction}

The analysis of absorption systems in the spectra of high-redshift quasars has represented
for several decades the only practical way to obtain information about the properties
of the intergalactic medium (IGM). The Ly$\alpha$ forest gives an unique
insight into the one-dimensional distribution of low density hydrogen along
the quasars' line of sight. Higher column-density features,
i.e. the Lyman-limit systems (LLSs) and the damped Ly$\alpha$ systems (DLAs), correspond
to denser concentrations of atomic hydrogen and, for this reason, can
be associated with galaxy formation.  Unfortunately the information from quasar absorption spectra is
almost always one-dimensional.

An interesting alternative to absorption studies is to try to detect 
the IGM in emission rather than in absorption.
The absorption of ionizing photons should be associated with the emission of 
fluorescent Ly$\alpha$ photons, as originally proposed by Hogan \& Weymann (1987). 
Detection of the fluorescent emission would provide a three-dimensional picture of the neutral
IGM and proto-galactic clouds at high redshift. 

In an earlier paper (Cantalupo et al. 2005, hereafter Paper I), we constructed models
of fluorescent emission in realistic hydrodynamical simulations. 
These simulations were normalized so as to produce the correct statistics for Lyman 
limit absorption systems.
For clouds that are optically thick to the ionizing radiation 
(i.e. Lyman limit systems and above), the surface brightness (SB) of
the fluorescent emission is set by the strength of the ionizing background.
Unfortunately, the intensity of the UV background at $z\sim3$ (e.g. Haardt \& Madau 1996)
corresponds to a maximum Ly$\alpha$ SB of $\sim3\times10^{-20}$\ergscmarcsec\ .
 Detecting such a low SB is very challenging 
with present-day instrumentation.
Blind searches in the field have only produced a number
of null results (Lowenthal et al. 1990; Mart\'inez-Gonzalez et al. 1995;
Bunker, Marleau \& Graham 1998).

The SB of fluorescent emission will be higher
if the local ionizing background is increased, e.g. for gas clouds
lying close to a bright quasar.  
In Paper I, we examined this effect and showed that it could boost the
fluorescent emission to detectable levels over a large volume, 
i.e. of order 30,000 comoving Mpc$^{3}$ for a sufficiently bright quasar.
The penalty for this boost is that the extra photo-ionization
increases the total hydrogen column density that is required to reach 
a given neutral column density, i.e. to produce the maximum fluorescent surface 
brightness. This is the analogue of the proximity effect seen in quasar absorption spectra. 
It means that maximally fluorescent sources near to a quasar will be systems that, 
in the absence of the quasar, would be DLAs rather than LLS.
DLAs have hydrogen column densities comparable to a sightline through the disk
of present day spiral galaxies and a total gas mass similar to the mass in
stars and gas within galaxies at the present epoch (Wolfe et al. 1995). For these reasons, DLAs
are widely believed to be the gas reservoirs from which the present galaxies formed.
However, it is still unclear if DLAs are galaxies already in place (e.g. Prochaska \& Wolfe 1998)
or are still in the form of protogalactic clouds (e.g. Haehnelt, Steinmetz \& Rauch 1998).
They typically contain very little 
of the molecular hydrogen usually associated with 
star formation (e.g. Ledoux et al. 2003, but see also Zwaan \& Prochaska 2006).
Moreover, metallicity measurements show a large range
from 1/10 solar to a value close to that  measured for the Ly$\alpha$ forest 
(cf. Prochaska et al. 2003 and Simcoe et al. 2004).  It is likely that DLAs
include systems of different origins.

In addition to studying the intergalactic medium, 
the detection and measurement of fluorescent Ly$\alpha$
emission from clouds at different distances and 
orientations from the quasar,
would in principle allow the study of the 
history and angular distribution of the ultraviolet emission from quasars. 
Both questions are of considerable topical interest in the context of unified models
of active galactic nuclei (see e.g. Antonucci 1993).

In this paper we describe a blind survey for fluorescent 
Ly$\alpha$ emission carried out
around the $z\sim3.1$ quasar QSO0420-388, one of the 
brightest quasars at high redshift for which the
Ly$\alpha$ line lies within an 
existing narrow-band filter of the VLT-FORS2 instrument. 

Previous searches for fluorescence around quasars have had either null results at $z\sim2.2$ 
(Francis \& Bland-Hawthorn 2004)
or only one (or marginally two) possible detection at $z\sim4.3$
(Francis \& McDonnell 2006).
In the first case, the null result is perfectly consistent with our 
model of fluorescent Ly$\alpha$ emission presented in Paper I. 
In the second case, the very close 
association of the system with the quasar itself
makes the identification with fluorescence difficult
(see section \ref{other_lit} for details).
Other possible fluorescent sources have been detected serendipitously in connection
with absorbing systems in quasar pairs (Fynbo et al. 1999, Adelberger et al. 2006).
However, the metallicity of the systems and the presence of a substantial continuum
emission probably indicate the presence of star formation within the clouds.
Ly$\alpha$ emission at high redshift can of course be produced by internally ionized
clouds, either from an associated young stellar population
or from processes involving the gas itself (e.g. cooling).
Several observations during the last decade have shown that these emitters  
are quite common in the high-redshift universe (for a review see Taniguchi et al. 2003)
and enhanced by clustering effects around bright sources like AGNs 
and, in particular, radio-galaxies (e.g. Venemans et al. 2005 and
references therein). In searching for fluorescence, we must thus take into 
account the 
presence of this population.

The layout of the paper is as follows. In \S 2 we compare the characteristics of 
fluorescent Ly$\alpha$ emission with those of other sources of Ly$\alpha$ emission at high redshift, in order
to determine which properties can best be used to recognize fluorescent emission. 
Although fluorescent sources should have characteristic equivalent widths (extremely high),
SB and spectral profiles, none of these provide a unique diagnostic.
  We conclude this section by
reviewing recent claims in the literature for detected fluorescence.
In \S 3 we describe our own new observations
and the data reduction. We then isolate a set of Ly$\alpha$ emitting candidates around the target quasar.
In \S 4, we discuss in detail the properties of the individual sources in the context of the expected
fluorescent characteristics, and compare their number density with
both our own models for fluorescence (Paper I) and those of other non-fluorescent
Ly$\alpha$ sources.  We conclude that it is likely, but not certain, that some of the detected sources
that we have detected are indeed fluorescent.  In \S 5 we summarize the paper.

\section{Recognition of fluorescent emission}

Besides fluorescent emission, Ly$\alpha$ can be produced by several kinds of
internal sources, e.g. photo-ionization from
a young stellar population or active nucleus (LAE, Ly$\alpha$ emitter [galaxies]),
or from cooling of shock-heated gas.
Many observations during the last decade have shown that Ly$\alpha$ emitters are
quite common in the high-redshift universe 
(e.g. Cowie \& Hu 1998; Pascarelle, Windhorst \& Kell 1998; Fynbo et al. 2003)
and enhanced by clustering
effects around objects like AGNs and, in particular, radio-galaxies 
(e.g. Le Fevre et al.1996; Pascarelle et al. 1996; Venemans et al. 2005) which likely 
reside in high density regions of the Universe.

In this section, we review the characteristic properties that in principle could identify 
a source as fluorescent: i) the equivalent width (EW), ii) the
emission profile, iii) the surface brightness (SB), and
iv) the number density of emitters.  We will take advantage of the model
recently presented in Paper I,
in which a combination of a high-resolution hydrodynamic simulation and two radiative
transfer schemes was used to predict the properties, spectra and number density of
fluorescent Ly$\alpha$ emitters at $z\sim3$.

\subsection{Equivalent Widths}

In the fluorescent case, the Ly$\alpha$ emission originates from pure gaseous clouds
externally photo-ionized. Therefore, any continuum emission comes only
from recombination processes: mainly two-photon continuum and free-bound recombinations. However, given the low value of the emission coefficient
corresponding to these processes,
the continuum level around the Ly$\alpha$ wavelength is very faint,
resulting in an extremely high EW. The case of Ly$\alpha$ emission originating
from cooling shock-heated gas will be similar. 

If the Ly$\alpha$ emission results from photo-ionization by
embedded young stars, we expect a significant contribution to the UV continuum emission 
from the stellar population. In this case, the Ly$\alpha$ EW
depends on the age, initial mass function (IMF), metallicity and dust content 
of the star-formation burst. Synthesis population models (see e.g. Schaerer 2003 and references
therein) show that high EWs can be produced for very young bursts in
metal-poor and dust-free galaxies. Upper limits for very metal-poor
models ($0<Z<10^{-7}$) range from EW$\sim800\mathrm{\AA}$ (for a very young
population with an age of 2 Myr) to EW$\sim200\mathrm{\AA}$ (for a 10 Myr
old star-formation burst). Synthesis models of star-forming
galaxies with moderate ages ($>10$ Myr) and with metallicity (and IMF) closer
to the normally observed values have EW in the range $50-200$
$\mathrm{\AA}$ (Charlot \& Fall 1993). Surveys of Ly$\alpha$ emitters at $z\sim3$
usually show a distribution of EW in the range from $15-100\mathrm{\AA}$,
with a few objects extending to higher values (up to 200-300$\mathrm{\AA}$).
High-EW objects can also be associated with AGN activity.

\subsection{Emission line profiles}\label{aelp}

The properties of the Ly$\alpha$ line have been widely
studied analytically under certain approximations
(e.g. Neufeld 1990 and references therein).
In particular, for an extremely opaque and plane-parallel medium,
it is possible to demonstrate that the emerging profile consists
of two symmetric peaks separated by a velocity shift of 
$\Delta v\sim2(\ln\tau_{\mathrm{Ly}\alpha})^{1/2}\sqrt{2}\sigma_{\mathrm{th}}$
(where $\tau_{\mathrm{Ly}\alpha}$ and $\sigma_{\mathrm{th}}$ are, respectively, 
the central-line optical depth of Ly$\alpha$ photons and
the thermal velocity dispersion of the medium).

 Since fluorescent photons originate from externally
ionized clouds, their typical $\tau_{\mathrm{Ly}\alpha}$ should correspond
to $\tau_{\mathrm{Ly}\alpha}\sim2\times10^4$ (assuming a temperature of
$T=2\times10^4$ K). Therefore, in the above approximations,
fluorescent emission should be characterized by two
symmetric peaks separated by $\Delta v\sim 8.9\sigma_{\mathrm{th}}$.

However, detailed simulations of fluorescent emission in more realistic 
situations show that the inclusion of velocity
fields inside and around the clouds influences
the emerging line profile. In most cases, the symmetry
of the double-humped profile is lost and one of the
two peaks is severely suppressed (see Fig. 7 in Paper I;
notice that the x-scale of that figure is erroneously
multiplied by a factor of 4).  Furthermore, if the gas
consists of several small clumps, a turbulent
velocity term should be added to the thermal velocity
dispersion, increasing the separation of the peaks.  
 On the other hand,  
if the turbulent velocity of the emitting region
is large compared with that of the intervening layers, the
symmetry of the line profile may be restored.  

While the double peaked nature of fluorescent emission can be
modified, the single Ly$\alpha$ lines generated
by internal photo-ionization at low optical depth can be partially
absorbed by infalling or expanding gas. As a result, the
emergent profile can again consist of multiple components. 
 
We conclude that these practicalities mean that the analysis 
of the emission line profile alone cannot 
provide a watertight signature of fluorescence. 

\begin{deluxetable*}{clccccc}[h!]
\tablecolumns{7}
\tablecaption{\label{obslog} Summary of the observations and setups.}
\tablehead{ Setup name & Date & Mode & Filter &\ \ Mask rotation\ \ &\ \ Combined seeing\ \ &\ \ Exposure time\ \ }
\startdata
S1  & Nov. 29 & MXU & OIII/3000+51 & 0$^\mathrm{o}$ & 1''.0 & 25 200 s \\
S2  & Nov. 30 and Dec. 1 & MXU & OIII/3000+51 & 90$^\mathrm{o}$ & 1''.0 & 34 200 s \\
IMA & Dec. 1 & Imaging & Bessel V & - & 1''.0 & 7 200 s \\
S3  & Dec. 2 & MXU & OIII+50 & 0$^\mathrm{o}$ & 0''.7 & 23 400 s \\
\enddata
\end{deluxetable*}

\subsection{Surface Brightness}

With a simplified slab geometry (e.g. Hogan \& Weymann 1987, Gould \& Weinberg 1996) of
a semi-infinite and static medium, 
fluorescent (self-shielded) clouds should act like ``mirrors": 
illuminated by an external source they return
the incident ionizing photons in the form of Ly$\alpha$ line emission
with an efficiency of $\sim65\%$.  In this case, the Ly$\alpha$ SB 
is uniform and its value is easily
predicted once the impinging ionizing flux is known, and vice-versa. 

In the case of a locally enhanced ionizing flux, e.g. from a nearby quasar,
the fluorescent SB would be expected in this simplified
geometry to be higher.  We introduce a 
``boost factor'' $b$, which is defined as the increase in the predicted SB 
relative to that
induced by the UV background at $z\sim3$ 
(${\mathrm{SB}}_{\mathrm{bg}}=3.67\times10^{-20}$\ergscmarcsec ), i.e. SB$=(1+b)$SB$_{\mathrm{bg}}$.
 The value of $b$ corresponds to:
\begin{equation}\label{bdef}
b=15.2 \,\frac{L_{\rm LL}}{10^{30}\, \mathrm{erg}\, \mathrm{s}^{-1} \,
\mathrm{Hz}^{-1}} \,
\frac{0.7}{\alpha}\,\left(\frac{r}{\mathrm{1\,Mpc}}\right)^{-2}
\end{equation}
at a physical distance $r$ from a quasar
with monochromatic luminosity $L_\nu=
L_{\rm LL}(\nu/\nu_{\rm LL})^{-\alpha}$
(see figs.6,9 and section 3.3 in Paper I for further details).

When velocity fields and a more realistic gas distribution are taken 
into account, it is found (Paper I) that the SB may be reduced.
In particular, for anisotropic illumination (i.e. fluorescence induced
by a quasar), 
the expected SB in the direction of the incoming flux
can be much lower than expected in the slab approximation. The importance of this effect,
which we called the ``geometric effect'', depends on the relative size of the shielding-layer
with respect to the cloud radius, and therefore on the gas density profile and the
intensity of the impinging radiation. In Paper I, we found that 
the fluorescent SB from self-shielded clouds is still 
predictable in good approximation from the empirical relation:
\begin{equation}\label{fSB}
\mathrm{SB}=(0.74+0.50\,b^{0.89})\mathrm{SB}_{\mathrm{bg}}\,
\end{equation}
where $b$ represents the ideal ``boost factor'' described above.

The SB predicted in this way represents a fundamental upper-limit constraint:
emitters with higher SB must receive a ionizing contribution from other sources
and, therefore, are with high probability internally ionized clouds.
However, several uncertainties remain to be taken into account. In particular,
the exact distance from the quasar may be uncertain due to peculiar motions (and
uncertainty in the precise systemic redshift of the quasar).  Anisotropy or
time-variability in the quasar emission will also modify the SB.
In fact this latter effect offers the possibility, if other indicators indicate
a fluorescent origin of the emission, that their SBes can be used 
to study the angular distribution and temporal history of the quasar UV emission.

\subsection{The number density of emitters}

The covering factor of fluorescent sources on the sky within a 
volume is directly related to the 
number density of optically thick absorbers seen in quasar spectra. This fact was used in 
Paper I to normalise our simulations.  Two complications enter in the present case.  
First, the regions around quasars are likely to have higher than average density.
Second, the enhanced ionizing radiation will produce a higher photo-ionization 
in a given cloud, reducing the column density of neutral Hydrogen.  The high surface
brightness fluorescence will still come from systems which would be seen as LLS
if there was a background quasar, but these are systems that would have been seen as DLA
if the extra ionizing radiation had not been present.  The fluorescence near to
a quasar is thus closely linked to the ``proximity effect'' (e.g. 
Bajtlik, Duncan \& Ostriker 1988).
 Since statistics of LLS near QSOs are not constrained, we must rely on theoretical
models to determine their covering factor.

The procedure to calculate the expected number of fluorescent 
emitters around a quasar, outlined in section 3.4 in Paper I, consists of three steps: 
i) convert the SB sensitivity limit of the survey (SB$_{\rm{lim}}$)
into a minimum boost factor $b_{\mathrm{min}}$ inverting Eq.(\ref{fSB}),
ii) use $b=b_{\mathrm{min}}$ in Figure 11 of Paper I to determine the physical number density of
fluorescent sources with SB$\geq$SB$_{\rm lim}$;
iii) given the values of $L_{\mathrm{LL}}$, $\alpha$ for the selected quasar, invert Eq. \ref{bdef}
using $b=b_{\mathrm{min}}$ to find the maximum distance ($r_{\mathrm{max}}$), and thus the volume, where
the fluorescent SB$\geq$SB$_{\rm lim}$.
As an example, a sensitivity of SB$_{\rm lim}\sim10^{-18}$\ergscmarcsec\ corresponds to
a fluorescent (physical) number density of $\mathrm{dN}/\mathrm{dV}\sim0.3\pm 0.1$ Mpc$^{-3}$
(for sources with area $A\simeq 4$ arcsec$^2$), neglecting any density enhancement around the
quasar.

This may be compared with the number density of other Ly$\alpha$ sources around 
$z\sim3$ quasars. Recent surveys indicate that the number of Ly$\alpha$ emitters
around radio-galaxies at that redshift can be enhanced by a factor $\sim3$
with respect to the field. In particular, Venemans et al. (2005; hereafter V05)
found an emitter number density of $0.26\pm0.04$ Mpc$^{-3}$ around the
$z\sim3.13$ radio-galaxy MRC0316-257, in a survey with sensitivity
of the order of SB$_{\rm lim}\sim10^{-18}$\ergscmarcsec\ . In the view of the
AGN unified model (see e.g. Antonucci 1993), the
environment of radio-loud quasars should be similar to that of radio-galaxies.
Moreover, we do not expect a significant contribution from fluorescent
sources to the number density found by V05, given the shape of their survey
and the direction of the (probable) UV emission-cone from MRC0316-257.

\subsection{Fluorescent candidates in the literature}\label{other_lit}

In this section, we briefly describe previous surveys for fluorescent emission and
review the results of these in the light of our discussion above. 
Given the difficulties in the detection of fluorescence
induced by the UV-background alone, previous surveys have centered around quasars, as in our case.

Francis \& Bland-Hawtorn (2004) reported an attempt to find fluorescence around a quasar
at redshift $\sim2.17$ basing their expectations on the simplified analytical models (Gould \& Weinberg 1996).
In particular, they were expecting to find at least 6 fluorescent emitters within their sampled
region of 1 Mpc around the quasar, but they saw none. However, this result is perfectly consistent
with the expectation of our more realistic model of fluorescent emission (see section 3.5 in Paper I
for details). 

Adelberger et al. (2006) reported the serendipitous discovery of a possible fluorescent emitter
associated with a DLA at redshift $\sim2.8$. In particular they found the
emitter in the spectrum of a Lyman-break object located 49'' from a much brighter quasar.
The object consists of two peaks with separation $\Delta V\sim500$ km s$^{-1}$ and its
SB seems to be compatible with the fluorescence induced by the quasar, even though the
required boost factor is extremely high ($b\sim2000$, in the ideal case of a slab). 
However, the object is clearly detected
in their $G$ and $R$ images (with an observed magnitude $G_{\rm AB}=26.8\pm0.2$) resulting in 
a rest-frame equivalent width EW$_0=72\pm20\mathrm{\AA}$.
Therefore, this candidate is more plausibly an internally ionized source.
A similar detection, associated with a lower redshift DLA system ($z\sim1.9$) in the spectrum of
a close quasar-pair, was reported by Fynbo et al. (1999). In this case a continuum counterpart
is not clearly detected, although the subtraction of the PSF of
the bright quasars makes this difficult. However, the extreme proximity of the quasar responsible of the 
possible fluorescent emission ($\sim20$kpc), indicates a plausible physical connection
between the two objects.

Finally, Francis \& McDonnell (2006) claimed the detection of a fluorescent object
located just 0.8'' from a quasar at $z=4.28$. Their object is spectrally and spatially
unresolved and
the flux is poorly constrained because of the proximity of the quasar.
There is a lower limit for the equivalent width of EW$_0>19\mathrm{\AA}$. Given the vicinity
of the quasar ($\sim5-50$ physical kpc) and the uncertainties in the flux and size measurement,
comparison with the theoretical SB is hard, although they note that the
predicted luminosity is $\sim400$ times greater than observed, if the object is fluorescent.
The sampled area around the quasar is very small ($5''\times7''$ in projection)
because of the Integral Field Spectrometer used, making any 
statistical comparison difficult.

\section{Observations and data analysis}

In this section we describe our own survey for fluorescent emission around 
QSO0420-388, including the data reduction and the selection of the Ly$\alpha$
candidates.

\subsection{VLT spectroscopy and imaging}

Observations were taken during four visitor-mode nights on the VLT 8.2 m telescope
Antu on November 28 - December 1, 2005. The FORS2 spectrograph 
was used in multi-object spectroscopy mode (MXU) at standard 
resolution ($0.25'\times0.25'\ \mathrm{arcsec^2\ pixel^{-1}}$)
to build  a ``multi-slit plus filter'' (Crampton \& Lilly 1999) configuration 
centered on the $z=3.1$ QSO [HB89]0420-388.

For all the observations, we used the 1400V grism giving a
 dispersion of 0.61 $\mathrm{\AA}$ pixel$^{-1}$,
in combination with the narrow band filters OIII+50 and OIII/3000+51.
The resolution
(R=2100, corresponding to $\sim150 \mathrm{km}\mathrm{s^{-1}}$) is large enough
to resolve the doublet [OII]$\lambda\lambda3726,3729$, which helps to distinguish
a Ly$\alpha$ emission from low redshift contaminants.
The use of the narrow band filters allowed us to put 14 slits 
(each 6.8' long and 2'' wide) parallel to each other and separated
by 25'' without a significant overlapping of the spectra on the CCD.
The total area covered by the slits in the mask corresponds to
3.17 arcmin$^2$.
The same mask was
used for all the observations. but with two different filters and two different 
spatial orientations: i) slits
parallel to North-South direction, ii) slits parallel to East-West direction.

The use of two different narrow band filters and two mask orientations
was chosen in order to increase the sampled volume in redshift and projected
space. For our setup, the central wavelength of the filters OIII+50 and OIII/3000+51 is, 
respectively 5001$\mathrm{\AA}$ and 5045$\mathrm{\AA}$ with a FWHM of 57$\mathrm{\AA}$ 
and 59$\mathrm{\AA}$, corresponding to a total redshift range $z=3.089-3.173$ 
for Ly$\alpha$.  Filter OIII/3000+51
was used for both the mask orientations (0$^\mathrm{o}$ and 90$^\mathrm{o}$
rotations) and filter OIII+50 for just one
orientation (0$^\mathrm{o}$ rotation). 

For each configuration, a series of several dithered exposures of 1800s each,
stepped along the slit by 5 arcsec,
were obtained for a
combined exposure time of 7 hours (filter OIII+51 and 0$^\mathrm{o}$ rotation), 9.5 hours
(filter OIII+50 and 0$^\mathrm{o}$ rotation) and 6.5 hours (filter OIII+51 and 90$^\mathrm{o}$ 
rotation).
On the third night we also obtained a V-band image of the same field for 
a total exposure time of 2 hours. It consisted of 120 dithered exposures
of 60 secs each, and it was obtained under photometric sky conditions. 
An overview of the observations and of the different configurations is summarized
in Table \ref{obslog}.

\subsection{Data reduction}

The spectroscopic and photometric data were reduced using standard 
packages (IRAF - Image Reduction and Analysis)
and additional IDL routines written by the authors. 

The data reduction consisted of several steps
including bias subtraction, spectral distortion correction,
flat-fielding (using ``sky-flats'' obtained from the
scientifical images themselves) and sky subtraction.
The individual sky-subtracted images were then combined
to obtain the three scientific images corresponding to the three different 
configurations (see Table \ref{obslog}). 
 The combination of the images was performed with an averaged sigma clipping
algorithm in order to minimize the effect of bad pixels and cosmic rays.
A pixel-by-pixel noise map associated with each science image
was generated from the statistics of the pixels in
the multiple image combinations.

Spectro-photometric calibration was based on
observations of different spectro-photometric
standard stars for each night of observation.
The magnitude zero-points derived from several standard stars
were consistent with each other within 0.02 magnitudes. 
A small Galactic extinction correction of $A_{\lambda}(V)=0.079$ 
mag (Schlegel et al. 1998) 
was applied.

The result of the data reduction was three final ``spectroscopic'' multi-slit images
(S1, S2 and S3; 
corresponding to the different configurations of mask orientation and filters
summarized in Table \ref{obslog}) plus the V-band image (IMA) of the same field.
The ensuing steps included: i) the detection of line emission in
S1, S2 and S3; ii) the identification of possible continuum counterparts of the line
emission in the V-band image; iii) the selection of probable Ly$\alpha$ emission
with a criterion based on the rest-frame equivalent width of the line.

\subsection{Line emission detection}\label{lem}

The detection of line emission candidates was performed with the help of the
software package SExtractor (Bertin \& Arnouts 1996). 
In order to make  
the detection process more efficient, we first removed from each image 
the regions corresponding to the brightest 
continuum spectra (typically stars and local bright galaxies).
Such regions were automatically identified 
for each slit by summing the flux over the spectral band and applying a sigma-clipping
algorithm to the resulting 1-dimensional array. We removed all the
regions where the integrated signal was greater than 3 times the standard deviation
for that  slit. The excluded regions were, respectively,
$\sim4\%$ and $\sim9\%$ of the total image for the 0$^\mathrm{o}$ and 90$^\mathrm{o}$ rotation configurations. 

The resulting images and corresponding noise-maps (obtained 
in the reduction process described above) were then processed by SExtractor.
In order to allow the most general search, we
varied the three most relevant parameters, i.e. the smoothing filter, 
the minimum detection threshold and the minimum detection area 
(i.e. the number of connected pixels). 
We used always standard filters (Gaussian, top hat) plus
our own filters  
with a shape designed to increase the S/N of the
double peaked emission.  The minimum detection threshold and the number
of connected pixel was varied, respectively, from 0.3 to 1.5, and from 3 to 30 pixels.
Each combination of the three detection parameters produced a separate catalog
of line emission candidates, for a total of 1344 catalogs for each dataset.  We
refer to these as ``positive'' catalogues.

We then applied exactly the same procedure on
the original images after they had been
multiplied by $-1$ to produce a negative image. Any emission objects detected in these 
``negative'' catalogues must of course be
spurious.  We therefore kept only those ``positive'' catalogs for which the
corresponding ``negative'' catalogue (generated with the same
SExtractor parameters) contained {\it no} detected sources.
 Notice that this procedure assumes Gaussian (rather than Poisson)
 noise. This is a good approximation since the noise is sky-dominated and 
the sky level is high.
  The final step consisted
of merging together the lists of objects in these surviving catalogues, removing the duplications.

The output we had from SExtractor, at this stage was
the central position and approximate size of the candidate line-emission for each 
of the three spectroscopic images.  These were used used as a first guess
of the photometric aperture in the calculation of the line and continuum flux (as
explained in the next sections).  


\subsection{Identification of possible continuum counterparts}

For each candidate detected in the spectroscopic images, we identified the corresponding
region in the V-band image and calculated the continuum flux $f_{\rm V}$.
The size of the photometric aperture was fixed at the slit width (2'') perpendicular to the
slit, and the size of the detected object in the spectrum, in the direction along the
slit.
The location of these apertures was obtained from accurate relative astrometry of the brightest 
continuum objects which
were detected both in the spectra and the V image.
The errors on $f_{\rm V}$ were estimated from integrating on random positions within the image. 
The resulting 1$\sigma$ limiting magnitude for a typical aperture of $2''\times2''$ was 
about 27.7 mag.

\subsection{Selection of Ly$\alpha$ candidates}\label{selec}

 In order to select candidate Ly$\alpha$ emitters we used a method
based on the line rest-frame equivalent width :
\begin{equation}
\mathrm{EW}_0=\frac{F_{\rm line}}{C_{\rm line}(1+z)}
\end{equation} 
where $F_{\rm line}$ is the flux of the emission line,
$C_{\rm line}$ is the continuum, if present, at the same wavelength of the line
and $z$ is the redshift of the emitter.
Since the value of $C_{\rm line}$ is too faint to be reliably measured from the 
spectra, we base it on the V-band flux density $f_{\rm V}$ assuming a power law 
continuum with slope $\beta$.
Including the contribution of the line-emission, the flux density 
measured in the V-band is, to first approximation: 

\begin{equation}\label{flux_den}
f_{\mathrm{V}}=\frac{F_{\rm line}\cdot\epsilon_{\mathrm{V}}(\lambda_{\rm line})}{\int \epsilon_{\mathrm{V}}(\lambda) \mathrm{d}\lambda}+
    C_{\rm line}\left[\frac{\lambda_{\mathrm{eff,V}}}{\lambda_{\rm line}}\right]^{\beta}\ ,
\end{equation}
where $\epsilon_{\mathrm{V}}$, and $\lambda_{\mathrm{eff,V}}$ are, 
respectively, the efficiency and the central
wavelength ($5561.9\mathrm{\AA}$) of the V-band filter. 
From Eq. (\ref{flux_den}) we derive:
\begin{equation}
\mathrm{EW}_0=\frac{F_{\rm line}}{f_{\mathrm{V}}-f_{\mathrm{line,V}}}
     \left(\frac{\lambda_{\rm eff,V}}{\lambda_{\rm line}}\right)^{\beta}
     (1+z)^{-1}
\end{equation}
where     
\begin{equation}
f_{\mathrm{line,V}}=\frac{F_{\rm line}\cdot\epsilon_{\mathrm{V}}(\lambda_{\rm line})}
     {\int \epsilon_{\mathrm{V}}(\lambda) \mathrm{d}\lambda}
\end{equation}
represents the equivalent flux density in the V image given by the line emission
alone.

The line flux ($F_{\rm line}$) is calculated within an
aperture on the spectrum (as explained in \S\ref{lem}).
Instead, the value of $\epsilon_{\mathrm{V}}(\lambda_{\rm line})$ is 
interpolated from the measured transmission curve of the V-band filter.
We have not tried to take into account the absorption of
the continuum bluewards of $\lambda_{\rm line}$, because Ly$\alpha$ always falls in the
blue wing of the V-band.
We cannot estimate the value of the slope ($\beta$) for each source, and so
a flat spectrum ($\beta=-2$) was assumed for all the objects. 
The actual value of $\beta$ has little effect on EW$_0$ - i.e. changing to
$\beta=-1$ (or $\beta=-3$) changes the EW by only $\sim10\%$.

In order to compare our results with the other surveys for Ly$\alpha$ emitters, we
selected objects with $EW_0>15\mathrm{\AA}$ as candidate Ly$\alpha$ line (see e.g.
Venemans et al. 2005 and reference therein).  For candidates with $f_{\mathrm{V}} < f_{\mathrm{V}}(1\sigma)$ 
we obtain a lower limit 
to EW$_0$ by assuming $f_{\mathrm{V}}= f_{\mathrm{V}}(1\sigma)$.

\begin{deluxetable}{cccc}
\tablecolumns{4}
\tablecaption{\label{ztable} Position and redshift of the Ly$\alpha$ candidates 
and the QSO.}
\tablehead{
Object & \multicolumn{2}{c}{Position} & z\tablenotemark{a} \\
{} & $\alpha_{\mathrm{J}2000}$ & $\delta_{\mathrm{J}2000}$ \\
}
\startdata
1 & \ \ 04:22:18.53\ \ \ & \ \ $-$38:45:15.6\ \ \ & \ \ 3.1197 \ \ \ \\
2 & 04:22:21.17 & $-$38:46:29.6 & 3.1047 \\
3 & 04:22:27.56 & $-$38:43:24.8 & 3.1087 \\
4 & 04:22:27.55 & $-$38:42:14.5 & 3.1599 \\
5 & 04:22:23.36 & $-$38:45:32.1 & 3.1538 \\
6 & 04:22:12.68 & $-$38:46:39.9 & 3.1634 \\
7 & 04:22:19.31 & $-$38:43:35.8 & 3.1298 \\
8 & 04:22:04.06 & $-$38:45:52.1 & 3.1187 \\
9 & 04:22:01.97 & $-$38:45:59.2 & 3.0896 \\
10 & 04:22:23.35 & $-$38:44:42.0 & 3.1263 \\
11 & 04:21:59.87 & $-$38:43:21.0 & 3.0936 \\
12 & 04:22:16.87 & $-$38:42:27.6 & 3.1127 \\
13 & 04:22:21.08 & $-$38:43:35.8 & 3.1298 \\
\hline
QSO & 04:22:14.76 & $-$38:44:50.7 & 3.110\tablenotemark{b} \\
    &             &               & 3.123\tablenotemark{c} \\
\enddata
\tablenotetext{a}{ typical error in the redshift estimation is 0.0015}
\tablenotetext{b}{ estimated by Osmer et al. 1994} 
\tablenotetext{c}{ estimated by Lanzetta et al. 1993} 
\end{deluxetable}

\subsection{Results}

Each candidate lying above the equivalent width threshold (assuming the redshift given
from the Ly$\alpha$ line) was visually inspected in order
to remove spurious objects like leftover cosmic rays.
This resulted in a list of 14 candidates, of which 12 are compatible to be 
Ly$\alpha$ lines while the remaining 2 show a double
peaked emission that can also be compatible with the [OII] doublet.
One of these two candidates has a $2\sigma$ detection in the V-band
image and was rejected as a probable OII emitter. The other one, labeled \#13,
does not show any continuum counterpart above $1\sigma$ (corresponding
to an equivalent width greater than 70$\mathrm{\AA}$ for OII) and, for this
reason is included in the Ly$\alpha$ sample.
One other OII emitter, for which the identification is considered
secure (see caption of Fig.\ref{Otwo} 
for more details), was found by visual inspection of the regions 
containing a bright continuum that were excluded from the
analysis at the beginning (see section \ref{lem}).  

The fraction of contaminants in our sample is thus $1/14\sim7\%$,
similar to the fraction ($\sim6.5\%$) of low redshift interlopers found in other surveys
of Ly$\alpha$ emitters at $z=3.1$ (e.g. Steidel et al. 2000; Venemans et al. 2005).
In the following section we will discuss in detail the main characteristics of
the 13 Ly$\alpha$ candidates and, in \S 5, their possible fluorescent origin.


\section{Properties of Ly$\alpha$ candidates}

\begin{figure*}
\plotone{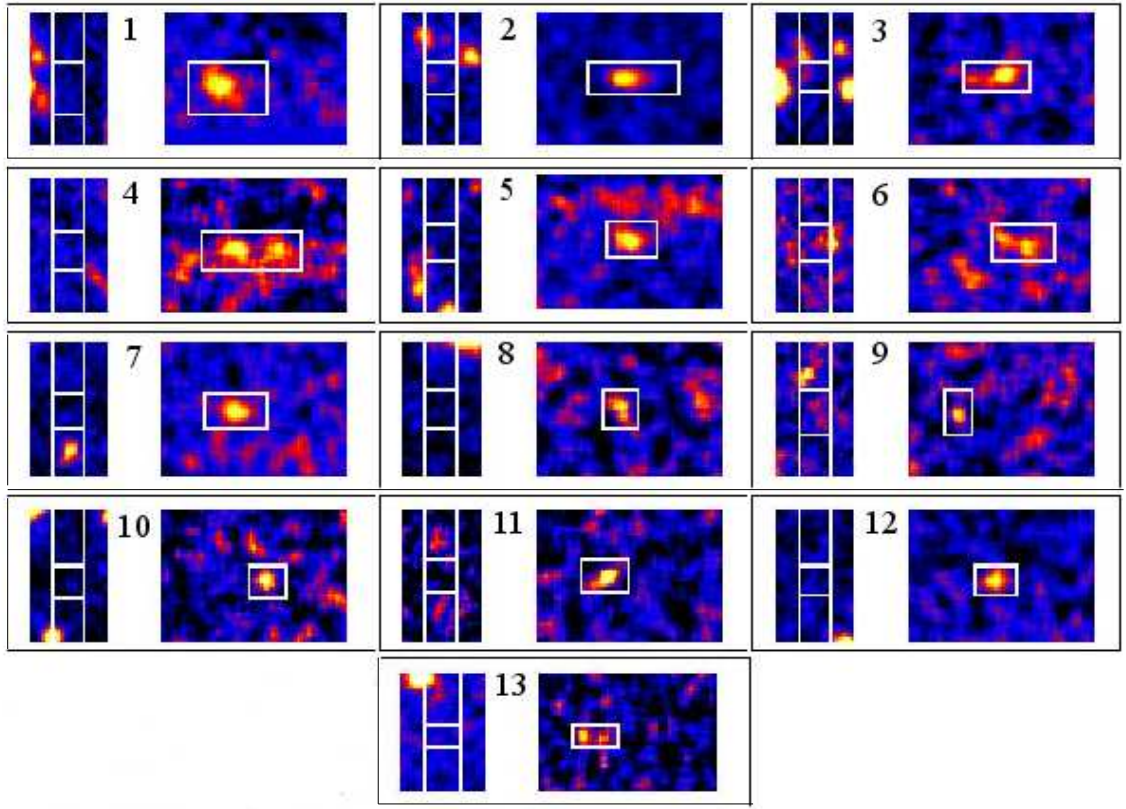}
\caption{2D-spectra (in S/N units) of the 13 Ly$\alpha$ candidates (right panels) and
the corresponding V-band image of the slit with the object positions (white boxes in
left panels). 
The dispersion direction in the spectra is parallel to the x-axis. The scale
of the panels is $32\mathrm{\AA}\times10''$ for the spectra and $4''\times10''$ for the
images. The white boxes represent the apertures used to calculate the line and V-band flux.
The values of flux and $1\sigma$ noise for each
aperture is shown in Table \ref{SPtable} (see text for details).}
\label{TwoDspc}
\end{figure*}

\begin{figure}
\plotone{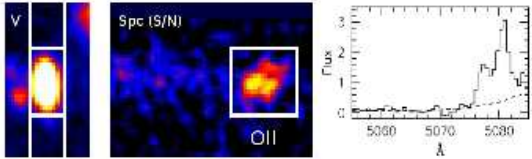}
\caption{V image (left panel), 2D-spectrum (in S/N units, central panel) and calibrated
1D-spectrum (solid histogram in right panel, dashed lines represent 1$\sigma$ errors) of a bright
OII emitter ($m_{\mathrm{V}}\sim17$) serendipitously found in one of our slits.
The emission wavelength, the distance between peaks ($\sim4\mathrm{\AA}$) and the flux 
ratio of the lines, all correspond to OII doublet
at $z\sim0.36$. The bright optical counterpart and the presence of a 
visible continuum redwards of the double emission line, 
also indicate that the emitter is a low-redshift galaxy.
} 
\label{Otwo}
\end{figure}

\begin{figure}
\plotone{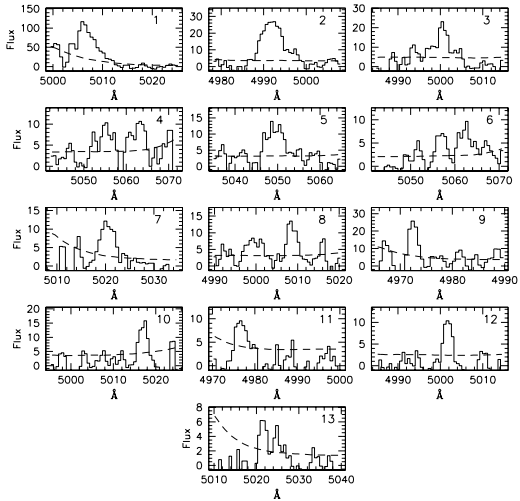}
\caption{1D calibrated spectra of the 13 Ly$\alpha$ candidates.
The flux (in units of $10^{-18}$\ergscmAA\ ) is obtained integrating within
the aperture represented by the white boxes in the left panels of Fig.\ref{TwoDspc}.
The dashed lines represent the 1$\sigma$ noise level.}
\label{OneDspc}
\end{figure}

The spatial distribution and the main spectro-photometric properties 
of the 13 Ly$\alpha$ candidates 
are summarized in Tables \ref{ztable} and \ref{SPtable}.
The 2-dimensional spectra in S/N units together with the corresponding
slit positions in the V-band image are shown in Fig. \ref{TwoDspc}, while the flux calibrated
1-dimensional spectra are shown in Fig. \ref{OneDspc}.
The photometric apertures used to calculate the line and continuum fluxes are
represented by the white boxes in Fig. \ref{TwoDspc}.

Only two objects (\#3 and \#6) have an integrated 
continuum flux in the V-band image that is well above $1\sigma$ for 
the corresponding photometric aperture.
However, in only one of these (\#6) there is a source clearly detected at the 
expected position.
In other cases (i.e. \#1, \#3 and \#5), the presence of 
close objects in the $V$-band image may have affected 
the photometry.
For the objects with $f_{\mathrm{V}}\lesssim1\sigma$, we estimate a lower limit for EW$_0$,
 assuming $f_{\mathrm{V}}=1\sigma$.  Thus, only one of
our candidates is clearly ruled out as a fluorescent emitter based on its EW$_0$. 

The spectral shape of the emission lines appears to be sharp 
and, in most of the cases, consists of only one clear peak.
One candidate (\#4) clearly shows the presence of two peaks separated by 
$\sim8\mathrm{\AA}$.  
Both peaks were independently detected by SExtractor at $S/N>5$. 
Candidates \#3 and \#6 also show a lower S/N ($\gtrsim3$) second peak
at a slightly different spatial position. Candidate \#13 was detected as a single emission line
but subsequent study of the unsmoothed
spectra revealed the presence of two emission lines separated by $3-4\mathrm{\AA}$. In other words,
about a third of the sample show some evidence for double line structure.

The estimation of the SB for Ly$\alpha$ candidates is complicated by their small sizes,
which typically range
between the seeing disk and 3 arcsec. 
For this reason, we have simply used the flux in the peak pixel, divided
by the product of the pixel and the slit size (i.e. $0''.25\times2''$) as a rough estimation
of the SB, that will, for the
unresolved objects, be a lower limit.
 Object \#4 shows a clear flat-topped profile over about $1".25$, and 
for this object we estimate the SB directly.

\subsection{Note on candidate \#1}

Candidate \#1 was detected in a part of the spectrogram in which the
spectra of two slits overlap. It was assigned to the blue edge of the appropriate
slit
because it lies closer to its central wavelength. We cannot completely rule out
the possibility that it belongs instead to the red edge of the adjacent slit. 
At the assumed position, candidate \#1 was situated close to a bright star that
was subtracted before calculating $f_{\rm V}$.

\begin{deluxetable*}{cccccl}
\tabletypesize{\scriptsize}
\tablecolumns{6}
\tablewidth{0pt}
\tablecaption{\label{SPtable} Spectro-photometrical properties of the Ly$\alpha$ candidates}
\tablehead{
Object & $F_{\rm line}\pm1\sigma$ & $f_{\rm V}\pm1\sigma$ & EW$_{0}$ & SB &\ \ \ 
Fluorescent?\tablenotemark{a} \\
{\#} & \ $10^{-18}$ \ergscm\ \  & \ $10^{-20}$ \ergscmAA\ \ \  & $\mathrm{\AA}$\ \ & \ \ $10^{-18}$ \ergscmarcsec\ \ \ &\ EW \ \ EP \ \ SB \ \\
}
\startdata
 1 &  58.27 $\pm$   6.67 &  6.55 $\pm$   5.90 & $\gtrsim$ 370 & 15.81 $\pm$ 2.73 &\plus \none \plus \\
 2 &  18.81 $\pm$   1.20 &  3.43 $\pm$   3.69 & $>$ 150 & $\gtrsim$ 9.72 &\plus \none \plus \tablenotemark{b} \\
 3 &  12.98 $\pm$   1.35 & 16.70 $\pm$   3.32 & 16$^{+6}_{-4}$ & 5.80 $\pm$ 1.06 &\minus \none \plus \\
 4 &  11.33 $\pm$   1.41 &  1.46 $\pm$   4.43 & $>$ 65 & 2.68 $\pm$ 0.78 & \plus \plus \minus \\
 5 &   7.30 $\pm$   0.81 &  4.76 $\pm$   4.06 & $\gtrsim$ 36 & 2.65 $\pm$ 0.49 & \none \none \minus \\
 6 &   6.36 $\pm$   0.68 &  8.86 $\pm$   4.06 & 15$^{+17}_{-5}$ & 1.93 $\pm$ 0.40 & \minus \none \minus \\
 7 &   5.31 $\pm$   0.82 &  0.86 $\pm$   4.06 & $>$30 & 2.71 $\pm$ 0.50 & \none \none \plus \\
 8 &   4.70 $\pm$   0.67 &  0.59 $\pm$   4.43 & $>$24 & 1.84 $\pm$ 0.49 & \none \none \plus \\
 9 &   4.21 $\pm$   1.16 &  0.65 $\pm$   5.17 & $>$18 & $\gtrsim$ 2.40 & \none \none \minus \\
10 &   4.06 $\pm$   1.01 & -5.65 $\pm$   3.69 & $>$25 & $\gtrsim$ 2.34 & \none \none \plus \\
11 &   3.49 $\pm$   1.05 & -1.98 $\pm$   3.69 & $>$21 & $\gtrsim$ 3.52 & \none \none \minus \\
12 &   2.65 $\pm$   0.56 &  3.17 $\pm$   3.32 & $>$18 & $\gtrsim$ 1.79 & \none \none \plus \\
13 &   2.62 $\pm$ 0.45 &  2.34 $\pm$ 2.58 & $>$23\ ($>$70\tablenotemark{c}) & 1.46 $\pm$ 0.39 & \none \plus \plus \\
 \hline
\multicolumn{5}{c}{Previous fluorescent Ly$\alpha$ candidates in literature} \\
\hline
AA1\tablenotemark{d}\ & 21.0$\pm$5.0  & $-$ & 72$^{+20}_{-20}$ & 110.00 $\pm$ 26.00 & \minus \plus \none \\
FD1\tablenotemark{e}\ & $\gtrsim$15.0 & $-$ & $>$19 & $-$  \\
\enddata
\tablenotetext{a}{in view of the indicators discussed in \S 2 
(where EP stays for emission-line profile)}
\tablenotetext{b}{if $z_{\rm QSO}=3.110$}
\tablenotetext{c}{if the line is OII at $z=0.36$}
\tablenotetext{d}{Adelberger et al. 2006, candidate at $z=2.842$}
\tablenotetext{e}{Francis \& McDonnell 2006, candidate at $z=4.279$}
\end{deluxetable*}


\begin{figure*}
\label{SBplot}
\plottwo{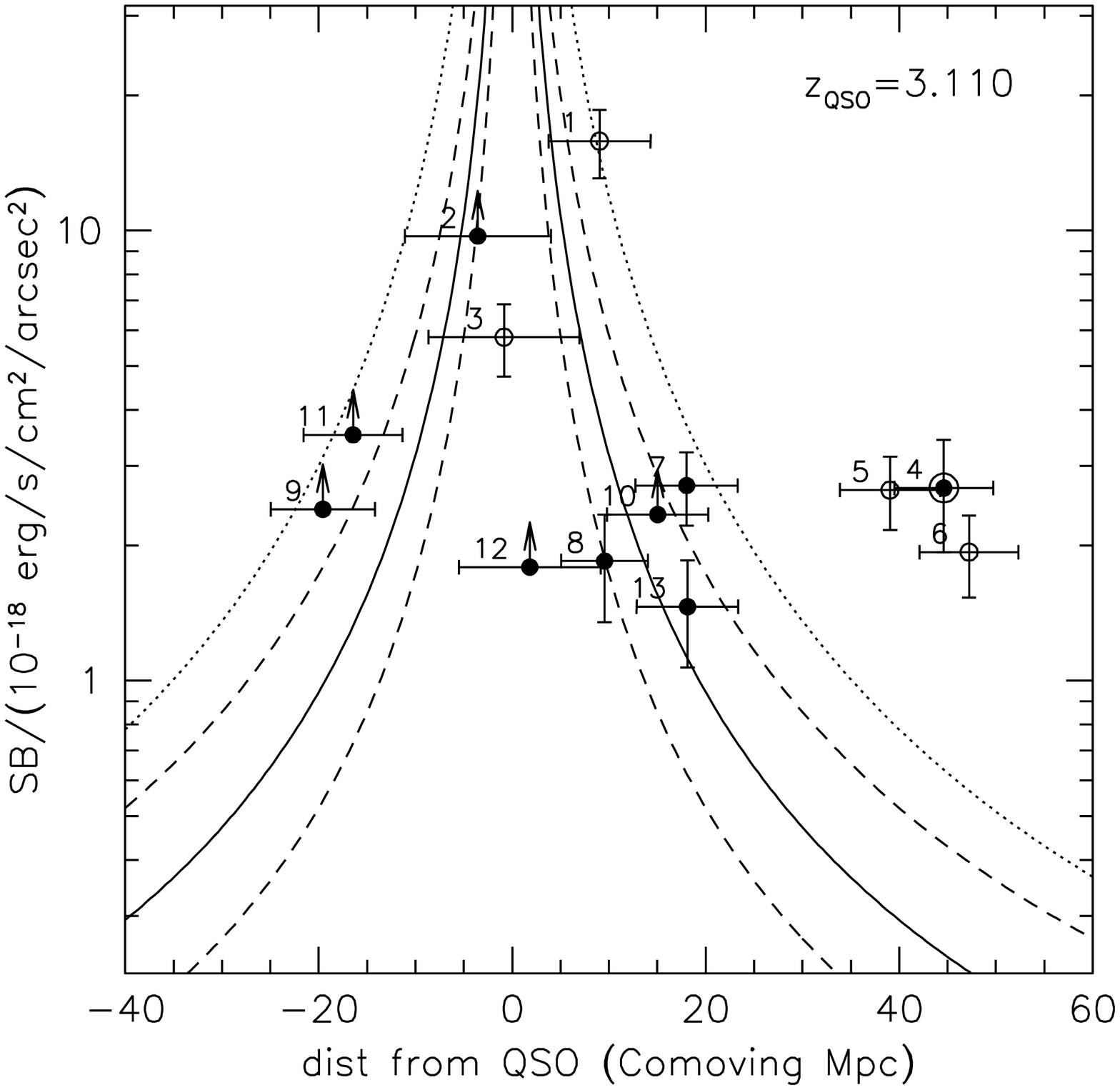}{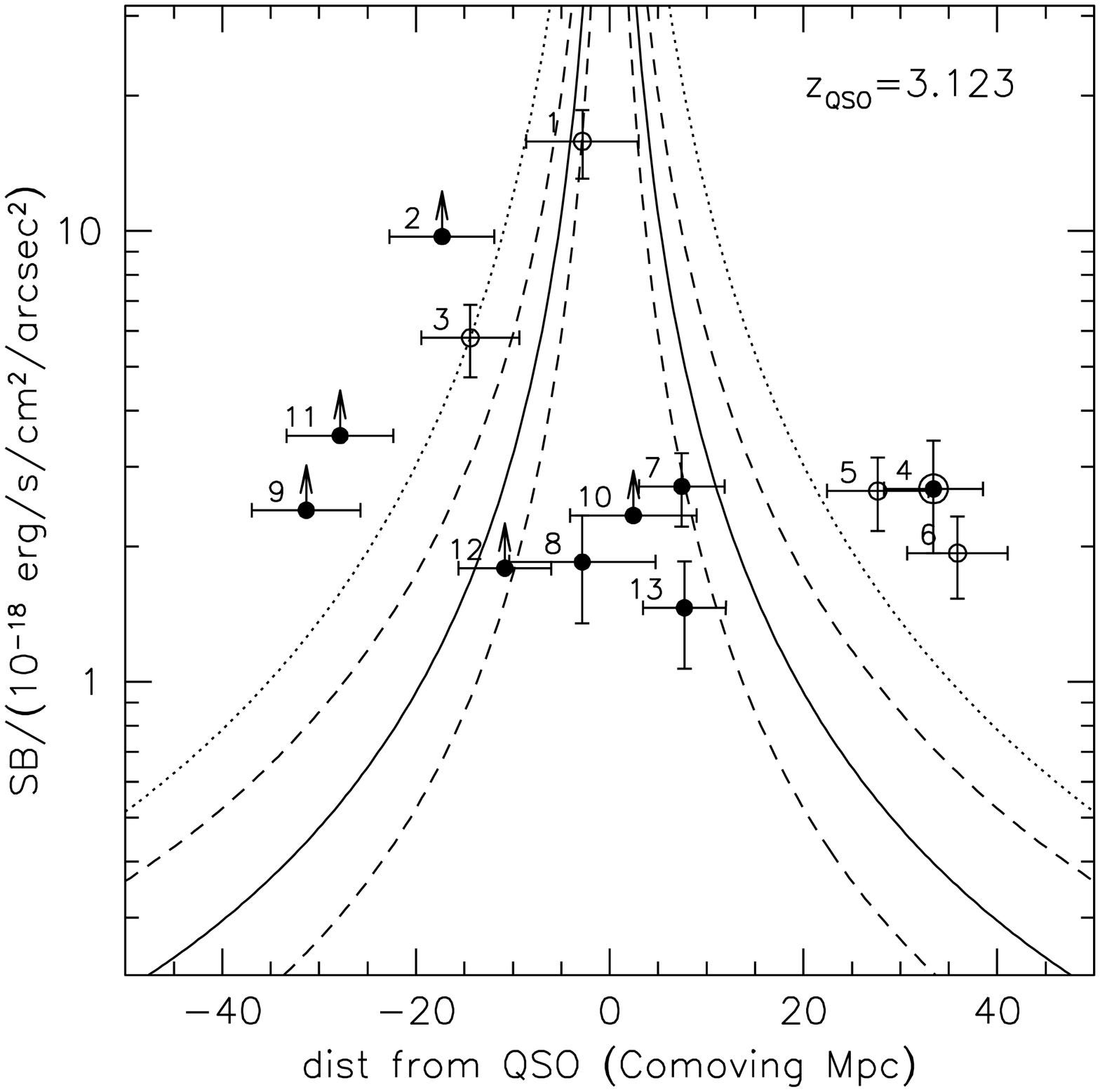}
\caption{Observed surface brightness (SB) of the 13 Ly$\alpha$ candidates against their three-dimensional
comoving distance from the
quasar for two different quasar redshifts.
The expected SB from self-shielded clouds, as calculated by the theoretical models
of Paper I, is shown by the solid line. 
The total errors in the predicted SB should take into account the scatter 
around the model relation and the uncertainties 
on the derived value of the quasar luminosity ($L_{\rm LL}$) and spectral shape ($\alpha$).
 To give an idea of the effect of these uncertainties in the predicted SB we overplot 
the expected fluorescent SB for a Lyman Limit flux 
of $2\times L_{\rm LL}$ (upper dashed line) and $0.5\times L_{\rm LL}$ (lower dashed line).
 The limiting case of fluorescent emission from a static slab is represented by the dotted line.
Open circles indicate the candidates with a V-band counterpart with flux greater
than 1$\sigma$ for the correspondent photometric aperture (see text for details).
The horizontal error bars take into account a peculiar velocity shift of $\pm250$km s$^{-1}$. 
All the candidates with a SB equal or lower than the theoretical expectations curves
are compatible with fluorescence induced by the quasar. 
}
\end{figure*}

\ \\

\section{Discussion}

Are any of our detected objects fluorescent? We know from \S 2 that a good
fluorescent candidate should have: i) very high EW, ii) possibly (but
not necessarily) a double-peaked profile, iii) a precise value of its SB.
We would also like to see the number density of objects to roughly
match the density calculated in \S 2, allowing for the uncertainty in clustering.

As a quick reference, we associated to each candidate in Table \ref{SPtable} a ``+''(or ``-'') if the
corresponding indicator clearly suggests (or disfavors) a fluorescent origin for the 
Ly$\alpha$ emission.
Regarding the first criterion we notice that,
given the sensitivity limits of our V-band image, 
only the two brightest emitters (\#1 and \#2) have high upper limits to their
EW (see Table \ref{SPtable}) - and it would be extremely difficult to
establish the required EW observationally.  The low
values of the EW for \#3 and \#6 indicate a likely non-fluorescent 
origin for the emission.

Two objects (\#4 and \#13) display a very clear signature of double-peaked emission.
Candidate \#4 is compatible with fluorescence if $\sigma_{\mathrm{th}}\sim60$ km s$^{-1}$.
Instead, object \#13 is compatible both with fluorescent Ly$\alpha$ 
(with $\sigma_{\mathrm{th}}\sim20$ km s$^{-1}$) and OII emission at $z=0.36$.
However, the high EW,
the intensity ratio of the peaks (i.e. that the blue peak is brighter than the
red peak) all disfavor the association with [OII] emission.

The comparison of the expected fluorescent SB
with the estimated SB of the individual candidates 
is shown in Fig. 4. Unfortunately the systemic redshift of the quasar is 
uncertain - $z=3.110$ is given by Osmer et al. 1994, and $z=3.123$ by Lanzetta et al. 1993.
The expected SB from self-shielded clouds (solid line) is calculated from Eq. (1), 
using the corresponding value of L$_{LL}$ ($\sim10^{32}\mathrm{erg}\mathrm{s^{-1}}\mathrm{Hz^{-1}}$) 
and $\alpha$ ($0.61\pm0.02$) for our quasar (extrapolated from Kuhn et al. 2001).
Only the most distant objects (4, 5 and 6) have an estimated SB that is clearly too 
high to be due
to fluorescence induced by the quasar, unless the quasar was brighter in the past. Note that two of these
(5 and 6) both have $f_{\mathrm{V}}\gtrsim1\sigma$ (and are  represented
by open-circle symbols in Fig. \ref{SBplot}), and 6 in particular has already
an EW that disfavors a fluorescent origin.  In general, we can notice that the case 
$z_{\rm QSO}=3.110$ allows a better fit of the expected 
relation between SB and distance for the detected objects.
In this case, at least $7-8$ candidates have a SB in good agreement with the theoretical
expectations.

The theoretical curves shown in Fig. \ref{SBplot} 
are calculated assuming isotropic and time-invariant emission
from the quasar.
The candidates beyond the quasar could have been
illuminated by a different UV flux with respect to the value measured by us.
In particular, gas clouds at 20 comoving Mpc from the quasar (within which most of our
candidates were found) should be responding to the ionizing flux of the 
quasar that was emitted or order $\sim3\times10^{7}$ yrs before the epoch of
observation. 
At 40 comoving Mpc, where candidates \#4, \#5 and \#6 lie, the time delay
is two times larger, i.e. $\sim6\times10^7$ yrs, close to the expected life-time
of the quasar (e.g. Porciani, Magliocchetti \& Norberg 2004).     

Turning the argument around, the effect of the time delay may represent a direct way to measure
the radiative history of quasars, if the fluorescent origin of the sources can be proved.
 If we assume that our object \#4 is truly fluorescent, which is plausible given the high EW and
the clear double-peaked profile, its SB would provide us with direct proof 
that the quasar was
shining $\sim10^8$ yr ago with an ultraviolet luminosity 6-10 times higher than the present measurements.
Confirmation of the fluorescent origin of object \#4 would for this reason be especially interesting.

In conclusion, what we can say about the fluorescent nature of our
detected objects? Unfortunately, there is no definitive evidence of fluorescence.
The EW can be poorly constrained for most of our candidates, 
requiring extremely deep broad-band images (for ground observations). The
SB-distance relation (as well as the emission-line profile), 
given all the theoretical and observational
uncertainties discussed above, is only an approximate indicator. 

Nevertheless, collecting all the indication so far, we can highlight 5 of the 13
objects which are most probably not fluorescent, i.e. \#3 (low-EW), \#5 (too high SB),
\#6 (both low-EW and too high SB), \#11 and \#12 (both objects are unresolved,
in front of the quasar and with SB at the limit for fluorescence).  However,
given their higher upper limit on the EW (mainly because of their high line-flux)
and their SB, objects \#1 and \#2 (if $z_{\rm QSO}=3.11$) could 
be good candidates for fluorescence.
As noted above, object \#4 could also be a good candidate, because the profile is flat in the center and
the presence of a double-peaked profile, 
compatible with fluorescence if $\sigma_{\mathrm{th}}\sim60$ km s$^{-1}$.  
Finally, objects \#7, \#8, \#10, \#12 and \#13 have SBs compatible
with fluorescent emission (for both quasar redshifts), but the EW is poorly constrained.

 As a final remark, it is worth stressing that self-shielded fluorescent
emitters around a quasar could also be detected in absorption
as the equivalent of DLAs in the field (see \S 1 for further details).
Therefore, the existence of such a
fluorescent population could in principle support 
the association of DLAs with proto-galactic gas clouds
(e.g. Haehnelt, Steinmetz \& Rauch 1998).
However, even if all fluorescent emitters around a quasar are DLAs, 
this does not imply the converse, allowing the possibility of a different origin
for a part of these absorbing systems, like disk galaxies already in place 
(e.g. Prochaska \& Wolfe 1998).

\subsection{The number density of emitters}

Given our sensitivity limits (SB$_{\rm lim}\sim10^{-18}$\ergscmarcsec\, for a typical aperture of 
$2''\times2''$) and the assumed value of $L_{\rm LL}$ and $\alpha$ for our quasar, we
can use the procedure discussed in \S 2 to derive the expected number of
fluorescent and not-fluorescent sources in our volume.
We find a physical number density of fluorescent emitters 
of $\mathrm{dN}/\mathrm{dV}\sim0.3\pm 0.1$ Mpc$^{-3}$, detectable within
$r_{\rm max}\sim 4.7$ physical Mpc from the quasar. Given the projected
area covered by our slits in all three configurations ($\sim 1.4$ physical Mpc$^2$), we therefore
should be able to detect fluorescent emission within a total volume $V\sim 13$ physical Mpc$^3$. 
The shape of this volume is approximatively a cuboid, being constrained from the limited field of view
of the detector ($6'.8\times6'.8$, corresponding to a physical Mpc scale of $3.2\times3.2$ 
at this redshift). 
Therefore the number of expected fluorescent emitters within $r_{\rm max}$ from the quasar
is $\sim4\pm 1$. Notice, however, that this number does not take into account: 
(i) the expected enhancement of the number density of objects around the quasar due to clustering effects, (ii) any reduction
due to beaming of the quasar emission - although we note that we would expect 
the illuminated cone to include the line of sight and thus to be more or less aligned with the
accessible volume. In this context we looked to see whether foreground and background 
candidates were located on different sides of the quasar, possibly indicating beaming,
but found no evidence of this.

By comparison, we expect
$\sim6$ non-fluorescent sources in our total sampled volume if it is 
similar to the radio galaxy environment studied by V05.

Despite the small statistics and with all the many uncertainties, these numbers agree very well with the
detected number of objects in our survey.  It may be a coincidence, but it is certainly consistent
with the indication that about half of our detected sources may have a fluorescent origin.

\section{Summary}\label{Concl}

Based on our recent model of fluorescent Ly$\alpha$ emission (Paper I),
we carried out a Ly$\alpha$ survey around the $z\sim3.1$ quasar QSO 0420-388 to search
for signatures of fluorescence. 
We used a ``multi-slit plus filter'' technique to sparsely sample a volume of
$\sim14000$ comoving Mpc$^3$ around the quasar (directly sampling $\sim1700$ comoving Mpc$^3$).

We found 13 emission line sources, selected on equivalent width, which are likely to be
Ly$\alpha$ at a redshift close to the quasar. 
In order to try to distinguish fluorescent objects from internally ionized clouds,
we measured three possible signatures of fluorescence: i) the line equivalent width, ii) the line
profile, iii) the SB. We also calculated the expected
number of fluorescent and non-fluorescent sources from theoretical models
and recent Ly$\alpha$ surveys.

From a theoretical point of view, the best constraints would be
a very high EW and, within some limitations, the relation of the SB
with the distance from the quasar.  Instead,
the line profile cannot give precise information on the nature of the sources - 
a double-peaked profile is a possible indication of fluorescence, but it can be absent, 
or be present in non-fluorescent sources.

Unfortunately, given the current observational limits, there is at
present no definitive evidence for fluorescence.
There are good reasons to suspect that 5 objects are probably
not-fluorescent, and this estimate is in good agreement with the number
density of such sources around radio galaxies. Of the remaining 8 objects, two or three of 
them are quite
good fluorescent candidates, based on moderately high upper limits to their EW,
and, in some cases, the presence of extended emission with a flat SB and
a double-peaked profile.  The remaining $5-6$ candidates are consistent with 
fluorescence, but their upper-limits on the EW are low enough to provide little
real constraint. 

On the basis of our model (Paper I), we were
expecting $3-5$ fluorescent sources in the inner region around the quasar,
and, from other surveys of Ly$\alpha$ emitters around AGNs, we were
expecting $\sim 6$ non-fluorescent sources in the total sampled volume.
These estimates are certainly consistent with the idea that about half of our detected 
sources could be fluorescent.

 In this case, the distance of most of the candidates from the ionizing source
 would imply that the quasar has been active for at least 30 Myr,
with a UV luminosity similar (within a factor of a few) to the present day value.
 In particular, one of the best candidates for fluorescence is sufficiently far behind 
the quasar to imply a life-time of $\gtrsim60$ Myr.
 
Although there are uncertainties in interpretation, due to both observational 
limitations and theoretical uncertainties (and no doubt a range of phenomena in Nature),
the current study provides one of the first statistical samples of possible
fluorescent emitters around a high-redshift quasar. 
Future studies on this category of objects will give information
about the properties of proto-galactic clouds and, furthermore, 
on the emission properties (i.e. spatial and temporal variations) of 
high-redshift quasars.

\acknowledgements
We are grateful to the staff of Paranal, Chile, for their support. 
S.C. thanks Claudia Scarlata for useful discussions and
acknowledges support from the Swiss National Science Foundation.

\end{document}